\documentclass[preprint,5p,twocolumn,final,number,sort&compress]{elsarticle}
\usepackage{amsmath} %
\usepackage{amsfonts}
\usepackage{amssymb} 
\usepackage{lineno,hyperref}
\modulolinenumbers[5]
\usepackage{graphicx}
\usepackage{textcomp}
\usepackage{xcolor}
\usepackage[capitalize]{cleveref}
\usepackage{comment}
\usepackage{natbib}
\usepackage{geometry}
\usepackage{txfonts}

\hypersetup{pdfauthor={Name}}

\journal{Nuclear Instruments and Methods in Physics Research Section A}

\biboptions{sort&compress}

\begin{document}

\begin{frontmatter}

\title{Exploration of Methods to Remove Implanted\\ \texorpdfstring{$^{210}$Pb}{Pb-210} and \texorpdfstring{$^{210}$Po}{Po-210} Contamination from Silicon Surfaces}

\author[1]{I.~J.~Arnquist}
\author[1]{R.~Bunker\corref{cor1}}
\author[2,3]{Z.~Dohnalek}
\author[1]{R.~Ma\fnref{fn1}}
\author[1]{N.~Uhnak}

\address[1]{Pacific Northwest National Laboratory, Richland, Washington 99352, USA}
\address[2]{Physical \& Computational Sciences Directorate and Institute for Integrated Catalysis, Pacific Northwest National Laboratory, Richland, Washington 99352, USA}
\address[3]{Voiland School of Chemical Engineering and Bioengineering, Washington State University, Pullman, Washington 99163, USA}
\cortext[cor1]{Corresponding author: raymond.bunker@pnnl.gov}
\fntext[fn1]{Now at Department of Anesthesiology, Weill Cornell Medicine, New York City, New York 10065, USA.}

\begin{abstract}
Radioactive contaminants on the surfaces of detector components can be a problematic source of background events for physics experiments searching for rare processes. Exposure to radon is a specific concern because it can result in the relatively long-lived $^{210}$Pb (and progeny) being implanted to significant subsurface depths such that removal is challenging. In this article we present results from a broad exploration of cleaning treatments to remove implanted $^{210}$Pb and $^{210}$Po contamination from silicon, which is an important material used in several rare-event searches. We demonstrate for the first time that heat treatments (``baking'') can effectively mitigate such surface contamination, with the results of a 1200 $^{\circ}$C bake consistent with perfect removal. We also report results using wet-chemistry and plasma-based methods, which show that etching can be highly effective provided the etch depth is sufficiently aggressive. Our survey of cleaning methods suggests consideration of multiple approaches during the different phases of detector construction 
to enable greater flexibility for efficient removal of  $^{210}$Pb and $^{210}$Po surface contamination.
\end{abstract}

\end{frontmatter}

\section{Introduction}
\label{sec:intro}
Radon is an important consideration for detectors that require low backgrounds from radioactivity to search for rare physics processes such as interactions from dark matter and neutrinoless double-beta decay (see, e.g., Refs.~\cite{Agnese:2016cpb,DarkSide-20k:2017zyg,LZ:2020fty,Benato:2017kdf,nEXO:2017nam,Cattadori:2021cnk}). Radon can manifest as a background in several ways. In this article, we focus on the long-lived surface contamination present on detector components exposed to radon, e.g., during fabrication and assembly which  often occur in environments with radon. Specifically, when $^{222}$Rn decays in air, its progeny plate onto surfaces where the relatively prompt alpha decays of $^{218}$Po and $^{214}$Po can implant $^{210}$Pb contamination to subsurface depths of many tens of nanometers~\cite{Stein:2017tel,ZUZEL2012140,Guiseppe:2017yah}. 
The long 22\,year half-life of $^{210}$Pb~\cite{NDS:Pb} and its decay chain support production of multiple types of background radiation---x-rays, betas, alphas, recoiling ions---for the full duration of detector operations. 

There are two primary mitigation approaches: use of radon-reduced environments to prevent radon exposure and surface cleaning to remove $^{210}$Pb and its progeny.  To achieve sufficiently low radon levels, the former can require significant infrastructure such as a dedicated low-radon cleanroom (as in Refs.~\cite{Street:2017bde,Grant:2011,doi:10.1063/1.2722065,Pocar:2005kp}). When practical, simple cleaning methods (such as wiping) are inexpensive but may be only partially effective because of the $^{210}$Pb implantation depth.  Exploration of more effective cleaning treatments is thus motivated by a need for cost-effective options for efficient removal of surface contaminants.

Methods for removal of radon-progeny surface contamination have been explored for a variety of materials and reported in the literature, including development of techniques to measure cleaning effectiveness in the low-radioactivity regime important for rare-event searches.  Cleaning of copper surfaces has been studied extensively, with methods ranging from relatively simple etching~\cite{HOPPE2007486,ZUZEL2012140,Zuzel:2018fzl,Bunker:2020sxw,NEWS-G:2020fhm} to more complicated multi-step procedures~\cite{ALESSANDRIA201313}. Other investigations have considered materials such as polymers~\cite{Stein:2017tel,Bruenner:2020arp,Broerman:2017agz}, stainless steel~\cite{ZUZEL2012140,Schnee:2013osi,Zuzel:2018fzl}, semiconductor crystals~\cite{ZUZEL2012149,Zuzel:2018fzl,Arnaboldi:2010fj,Street:2020xvb}, and glass~\cite{Kobayashi:2015lla}, using a variety of cleaning techniques such as leaching, etching, and electropolishing. Notably, silicon is an important detector medium used in several rare-event experiments~\cite{Agnese:2016cpb,DAMIC:2020cut,SENSEI:2020dpa,Aguilar-Arevalo:2022kqd,CDMS:2013juh,SuperCDMS:2020ymb,SuperCDMS:2020aus}.  There are many reports on surface treatments to advance fabrication of silicon-based devices (e.g., as in Refs.~\cite{HU2019268,pssa.201700152,WOS:000248830500016,WOS:000168449500041,WOS:000078232800056}). However, there appear to be few studies focused on removal of radioactive contaminants from silicon surfaces, such as the one in Ref.~\cite{Street:2020xvb} which specifically targeted crystal sidewalls.

\begin{figure*}[t!]
\centering
\includegraphics[trim=0 0 0 0, clip,width=0.99\textwidth]{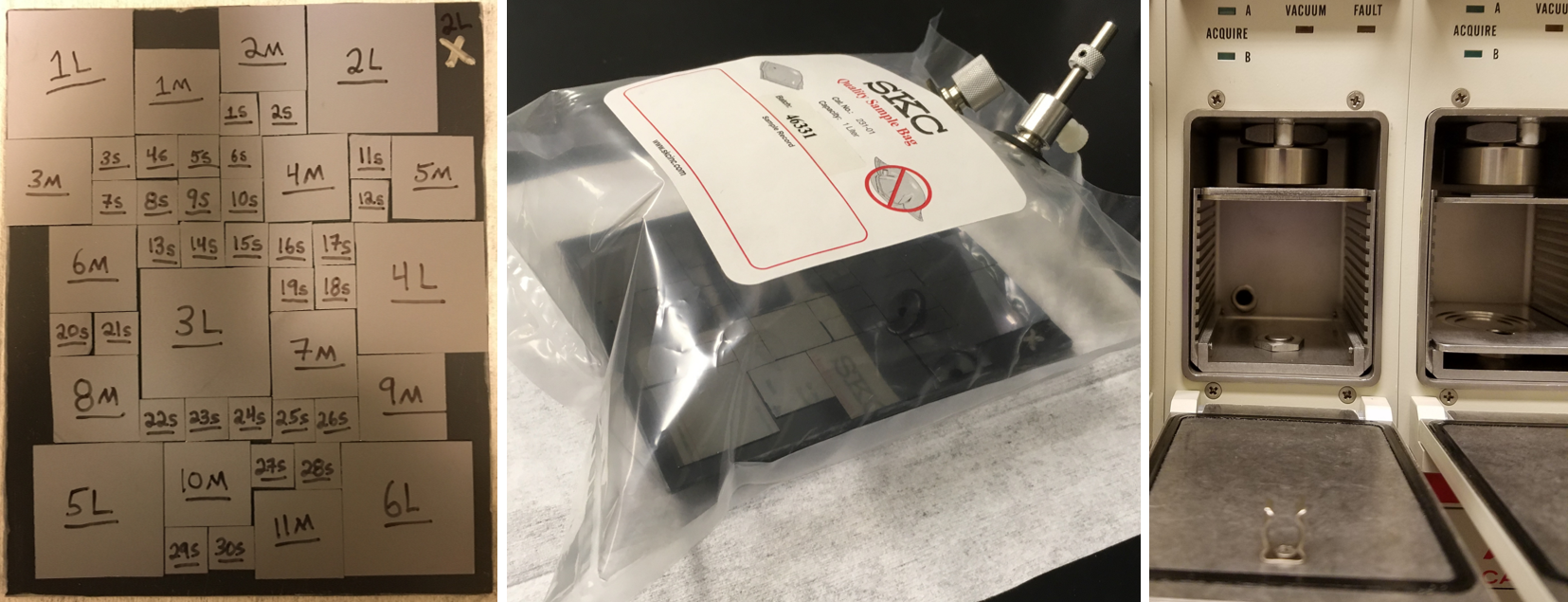}
\caption{\textit{Left:} Silicon coupons arranged on PVC backing plate, showing labels written on bottom. \textit{Middle:} Coupons affixed face up on backing plate and heat sealed inside a low-permeability Tedlar\textsuperscript{\textregistered}-film gas-sample bag, which is equipped with gas fittings and filled with nitrogen. \textit{Right:} Photo of 1$\times$1\,cm$^2$ (left) and 3$\times$3\,cm$^2$ (right) coupons as loaded inside alpha counting chambers, with the sample shelves placed high such that the coupons are located directly below the detectors at the top of the chambers to maximize detection efficiency.}
\label{fig:prep}
\end{figure*}

In this article, we report results to remove implanted $^{210}$Pb and $^{210}$Po from silicon surfaces using a variety of methods. Recognizing that there may be opportunities for surface treatments during different stages of detector fabrication (for other materials as well as silicon), we performed a general exploration using a broad range of processes, including heat treatments, wet chemistry, and plasma etching. The objective is to identify a suite of cleaning options to enable greater flexibility to effectively mitigate radon-progeny surface contamination. The paper is organized as follows. Section~\ref{sec:methods} describes our overall approach, including our preparation of $^{210}$Pb-implanted coupons in Sec.~\ref{ssec:samples} and our surface-contamination measurement and analysis methods in Sec.~\ref{ssec:assay}. Section~\ref{sec:results} motivates the different surface treatments, describes the specific procedures that we used, and presents our $^{210}$Pb and $^{210}$Po removal results. Finally, we summarize and discuss our results in Sec.~\ref{sec:summary}.

\section{Methods}  
\label{sec:methods}

\subsection{Sample Preparation}
\label{ssec:samples}
As shown in Fig.~\ref{fig:prep}, we prepared an assortment of silicon coupons from a pair of  single-side polished, FZ-grown ($>$10\,k$\Omega$-cm resistivity) wafers, each originally 100\,mm in diameter and 500\,$\mu$m thick.  The square-shaped coupons were diced `by hand' to provide three approximate sizes:  1$\times$1, 2$\times$2, and 3$\times$3\,cm$^{2}$.  This range of sizes enabled flexibility to explore a variety of surface treatment methods, while also providing good detection efficiency for $^{210}$Po surface contamination using conventional silicon alpha counters. From coupon to coupon, there are some minor differences in size as a result of the hand dicing, but they are generally not important because our individual results are based on relative rates measured before and after application of a surface treatment to a single coupon.

The coupons were packaged in a gas-sample bag for exposure to radon and thus implantation of $^{210}$Pb surface contamination. We arranged the full set on a PVC backing plate and labeled the unpolished side of each coupon (see Fig.~\ref{fig:prep} left).  For the radon exposure, each coupon's polished side was placed face up, with a small square of double stick tape under a corner to hold it in place. The backing plate and affixed coupons were then heat sealed inside a low-permeability Tedlar\textsuperscript{\textregistered}-film bag equipped with stainless-steel gas fittings.\footnote{1 liter Tedlar\textsuperscript{\textregistered} sample bag from SKC (part number 231-01A).} The bag was flushed with nitrogen gas---sequentially inflated and deflated three times---and finally inflated with nitrogen prior to injecting a radon sample via the septum gas fitting.  The middle panel in Fig.~\ref{fig:prep} shows the fully inflated sample bag during the flushing sequence.

While it is desirable to have a large activity to measure the effectiveness of our of methods with good statistical precision, we recognized that it would be impractical to introduce levels of unsealed radioactive material requiring additional controls into the laboratory spaces planned for sample processing. For the radon exposure, we therefore targeted the maximum activity allowed that would not impose additional controls for high levels of unsealed radioactivity. For our lab this limit is less than 20 alpha decays per minute per 100 cm$^2$, which corresponds to a $^{210}$Pb surface activity of a few mBq/cm$^2$. Despite our coupons' relatively low resulting $^{210}$Po activities, pre-treatment alpha count rates were nevertheless one to two orders of magnitude larger than the alpha counter background, depending on sample size and time since the radon exposure. 


The coupons were exposed to $^{222}$Rn extracted from a Pylon model RN-1025 source. The source can generate a total radon activity of 729\,kBq within its 106\,cc volume, which is more activity and more gas volume than needed for our exposure. For a 1\,week exposure, we estimate the $^{210}$Pb surface activity
\begin{equation}
    A_{\mathrm{Pb}} \approx \frac{f}{SA \times \tau_{\mathrm{Pb}}}  \int^{\mathrm{7 days}}_{0} \left(A_{\mathrm{Rn}} e^{\frac{-t}{\tau_{\mathrm{Rn}}}} dt\right)\mathrm{,}
\end{equation}
where $SA \approx 128$\,cm$^{2}$ is the total surface area of the coupons, $f$ is the fraction of radon decays for which the progeny plate onto the coupons, $A_{\mathrm{Rn}}$ is the total radon activity extracted from the Pylon source, and $\tau_{\mathrm{Rn}} = 5.5$\,d~\cite{NDS:Rn} and $\tau_{\mathrm{Pb}} = 32$\,yr~\cite{NDS:Pb} are the $^{222}$Rn and $^{210}$Pb mean lifetimes, respectively. We estimate $f \approx 20$\%, corresponding to the fractional surface-area coverage of the coupons inside the exposure bag.  To achieve a target $^{210}$Pb activity of a few mBq/cm$^2$ therefore suggests $A_{\mathrm{Rn}}$ of $\mathcal{O}(10\mathrm{\,kBq})$. 

The output of the Pylon source was equipped with a tee, with a valve attached to one outlet to enable flow through and a septum attached to the other outlet to enable extraction of radon with a syringe.  The Pylon source was reset by flushing it with nitrogen, and then all valves were closed to allow the $^{222}$Rn to build to full strength over a several-week period. 2\,cc of radon-spiked nitrogen was extracted from the source and subsequently injected into the Tedlar\textsuperscript{\textregistered} bag. Unfortunately, although the bag was tested to be gas tight prior to the radon injection, during the radon exposure the bag partially deflated.  It seems likely that the septum developed a leak as a result of the injection, potentially indicating the need for a finer-gauge syringe needle. Because of the weight of the bag's stainless-steel gas fittings, the bag deflated more at the end corresponding to the top of the backing plate (see Fig.~\ref{fig:prep} left and middle). The resulting surface activities exhibit some nonuniformity across the plate, with coupons toward the top of the plate less active than those toward the bottom; e.g., coupon 6L has $\sim$3$\times$ higher activity than coupon 1L. The coupons' $^{210}$Pb surface activities are generally less than 1\,mBq/cm$^{2}$. 

\subsection{Measurement and Analysis}
\label{ssec:assay}

\begin{figure}[t!]
\centering
\includegraphics[width=0.99\linewidth]{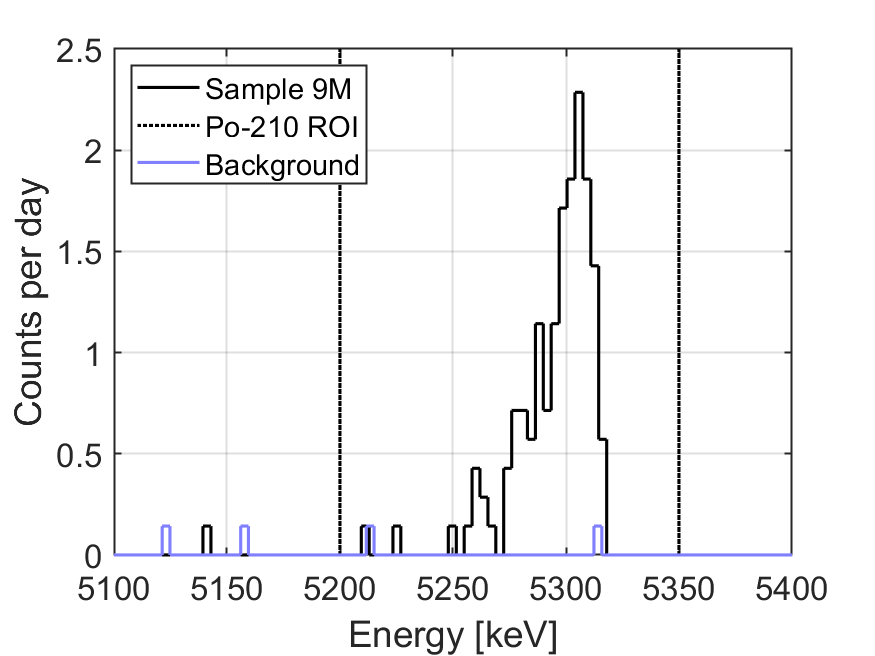}
\caption{Example alpha spectrum (black)---measured for Si coupon 9M---showing a clear peak at the  energy expected for $^{210}$Po alphas (5304\,keV~\cite{Kondev:2008roc}), compared to a typical background spectrum (blue) and the region of interest (ROI) used to measure the $^{210}$Po count rate (black dotted). }
\label{fig:spectrum}
\end{figure}

Coupons were counted using a set of Canberra alpha counters\footnote{Canberra model A450-18 PIPS\textsuperscript{\textregistered} (Passivated Implanted Planar Silicon) detectors housed in a Canberra Alpha Analyst 7200, Dual Spectrometer Module.} (see Fig.~\ref{fig:prep} right). Figure~\ref{fig:spectrum} shows a pre-treatment alpha spectrum compared to a typical background measurement. We use a relatively large region of interest (ROI) for measuring the $^{210}$Po alpha count rate to ensure a consistently high selection efficiency for the $^{210}$Po alphas while still maintaining a relatively low background. To maximize detection efficiency, coupons were placed as close to the face of the detector as practical. For the smaller 1$\times$1 and 2$\times$2\,cm$^{2}$ coupons, a second sample shelf with circular gradations was used to center coupons below the detector, whereas the larger 3$\times$3\,cm$^{2}$ coupons were centered on a shelf with a smooth surface placed in the topmost shelf position (see Fig.~\ref{fig:prep} right). For these counting configurations, we expect high detection efficiency for $^{210}$Po decays---a few tens of percent depending on coupon size and shelf placement. The exact efficiency is not important for our results, which are based on relative count rates before and after application of treatment methods.  Consequently, we did not  quantify the detection efficiency; rates are quoted in terms of the uncorrected event rate in counts per day (as in Fig.~\ref{fig:spectrum}).

Unexposed silicon coupons were measured to assess alpha counting backgrounds for the different coupon sizes, yielding 0--4 events in the $^{210}$Po ROI for a 1-week counting period.  This background level depends on the size of the coupon, because the silicon tends to shield the detector from the stainless-steel shelf background. Based on our measurements, we modeled the background in the $^{210}$Po ROI as a constant rate of 1, 2 or 3 counts per week for the 3$\times$3, 2$\times$2 and 1$\times$1\,cm$^{2}$ coupons, respectively. 68.3\% Feldman-Cousins unified confidence intervals are used to estimate statistical uncertainties~\cite{FC}.  In the low-statistics limit, these asymmetric uncertainties avoid unphysical confidence intervals (e.g., negative count rates).

To assess pre- and post-treatment $^{210}$Pb and $^{210}$Po surface activities, we measured the alpha count rate as a function of time. In general, we made two to three 1-week measurements over a span of several months---both before and after application of the treatment---so that the $^{210}$Pb level could be estimated from a time-dependent fit to the $^{210}$Po alpha count rate.

The pre-treatment $^{210}$Po count rate was modeled assuming zero $^{210}$Po and $^{210}$Pb activity prior to the radon exposure:
\begin{equation}\label{eq:PoPre}
    A_{\mathrm{Po}}\left(t\right) = A_{\mathrm{Pb}}^{0}\left[1-e^{\frac{-t}{\tau_{\mathrm{Bi}}+\tau_{\mathrm{Po}}}}\right]e^{\frac{-t}{\tau_{\mathrm{Pb}}}}H\left(t_c-t\right)\mathrm{,}
\end{equation}
where $t$ is the time since the radon exposure, $A_{\mathrm{Pb}}^{0}$ corresponds to the $^{210}$Pb activity directly following the radon exposure (at $t = 0$) and is a free parameter to be fit, $t_c$ is the time of the surface cleaning, $H$ is the Heaviside step function, and 
$\tau_{\mathrm{Bi}} = 7.2$\,d and $\tau_{\mathrm{Po}} = 199.6$\,d are the $^{210}$Bi and $^{210}$Po mean lifetimes, respectively~\cite{NDS:Pb}. 

\begin{figure*}[t!]
\centering
\includegraphics[width=0.475\linewidth]{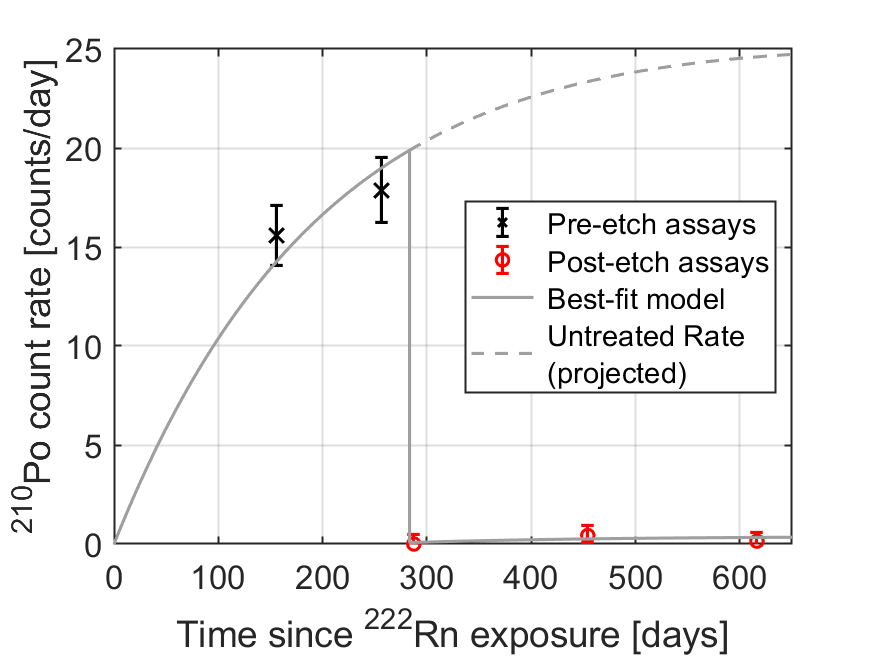}
\includegraphics[width=0.475\linewidth]{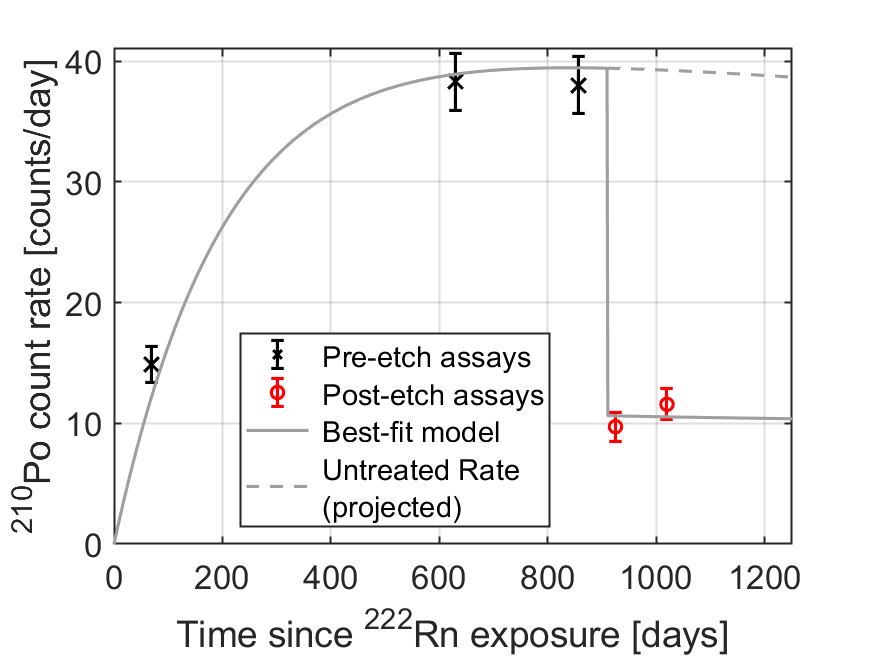}
\caption{Measured $^{210}$Po count rates for Si coupons 6M (left) and 6L (right) before (black $\times$) and after (red $\circ$) application of a wet etch using HF (see Sec.~\ref{ssec:chemical}). The $^{210}$Po in-growth that resulted from the radon exposure is apparent, and there is a clear reduction of the surface contamination following the wet etch. The reduction level  correlates with the strength of the HF solution applied to the coupon (see Fig.~\ref{fig:etch}). In each case, a combined fit of the models in Eqs.~\ref{eq:PoPre} and \ref{eq:PoPost} (solid gray) agrees well with the measured rates. For comparison, the pre-treatment $^{210}$Po count rate is projected forward in time (dashed gray) based on the best-fit value of $A_{\mathrm{Pb}}^{0}$. }.
\label{fig:ingrowth}
\end{figure*}

Where the data permit, both the $^{210}$Pb and $^{210}$Po surface-activity reductions, $f_{\mathrm{Pb}}$ and $f_{\mathrm{Po}}$, are determined by modeling the post-treatment $^{210}$Po count rate as
\begin{equation}\label{eq:PoPost}
\begin{split}
    A_{\widetilde{\mathrm{Po}}}\left(t\right)=
    A_{\mathrm{Pb}}^{0}\bigg[&\left(1-f_{\mathrm{Po}}\right)e^{\frac{-t_c}{\tau_{\mathrm{Pb}}}}\left(1-e^{\frac{-t_c}{\tau_{\mathrm{Bi}}+\tau_{\mathrm{Po}}}}\right)\,e^{\frac{t_c-t}{\tau_{\mathrm{Po}}}} +\\
    &\left(1-f_{\mathrm{Pb}}\right)e^{\frac{-t}{\tau_{\mathrm{Pb}}}}\left(1-e^{\frac{t_c-t}{\tau_{\mathrm{Bi}}+\tau_{\mathrm{Po}}}}\right)\bigg]H\left(t-t_c\right)\mathrm{,}
\end{split}
\end{equation}
where  $f_{\mathrm{Pb}}$ and $f_{\mathrm{Po}}$ are free parameters to be fit. If the data are insufficient to differentiate the post-treatment $^{210}$Pb and $^{210}$Po activities, we assume a common reduction factor such that Eq.~\ref{eq:PoPost} has one less free parameter.

Figure~\ref{fig:ingrowth} shows examples of the best-fit models for two coupons treated with different HF etching recipes (see Sec.~\ref{ssec:chemical}). In both cases, a simultaneous fit is performed to the measured pre- and post-treatment count rates versus time by summing Eqs.~\ref{eq:PoPre} and \ref{eq:PoPost} into a single, combined model. The best-fit model provides an estimate of either both $f_{\mathrm{Pb}}$ and $f_{\mathrm{Po}}$ (as in Fig.~\ref{fig:ingrowth} left) or a common reduction factor $f \equiv f_{\mathrm{Pb}} = f_{\mathrm{Po}}$ (as in Fig.~\ref{fig:ingrowth} right). The results discussed in the next section and plotted in Figs.~\ref{fig:heat}, \ref{fig:leach}, and \ref{fig:etch} are in terms of these best-fit reduction factors. Also shown in Fig.~\ref{fig:ingrowth}, the best-fit estimates of $A_{\mathrm{Pb}}^{0}$ are used to project the pre-treatment $^{210}$Po count rates forward in time ($t > t_c$) for comparison to the best-fit models.

\section{Surface Treatments and Results}
\label{sec:results}

\subsection{Heat Treatments}
\label{ssec:heat}

The use of heat treatments (or ``baking'') as a low-background method for removal of radioactive contaminants is generally unexplored in the literature. Studies of hydrogen diffusion in germanium and silicon~\cite{WOS:A1960WH86900023,WOS:A1987H899600001,WOS:A1968B254100003,WOS:A1982NF33100162} suggest that baking may be an effective mitigation technique for tritium. Also considering the high melting point of silicon relative to Pb, Bi, and Po ($>$1000\,$^{\circ}$C difference) motivated us to investigate baking as a cleaning method for our coupons. The potential for removal of $^{210}$Pb surface contamination is interesting because substrates are sometimes elevated in temperature during device fabrication; as a result, surface contamination from prior exposure to radon may be conveniently reduced. Our goal with this study was to experimentally demonstrate the plausibility of thermal treatments for removal of radon progeny from silicon surfaces. 

We applied baking protocols to three of our 3$\times$3\,cm$^2$ coupons.  Coupons 2L and 3L were each baked in a programmable TransTemp  quartz tube furnace (Thermcraft Inc., Winston-Salem, NC) set to ramp up to 1000\,$^{\circ}$C in 10\,min and hold this temperature for 12\,hr. The furnace's temperature monitor recorded a maximum temperature of 950\,$^{\circ}$C, reached after 20--30\,min and then held for the remainder of the 12\,hr bake. During each bake, the quartz tube was setup with a 2.6\,L/min flow of cover gas, where air was used for coupon 2L and boil-off nitrogen was used for 3L. These first two bakes resulted in partial  removal of the $^{210}$Pb and $^{210}$Po surface contamination, which motivated us to test if a much more aggressive baking protocol could yield perfect (or near-perfect) removal. Coupon 1L was baked in a Deltech DT-31 vertical muffle furnace at 1200\,$^{\circ}$C in a static air atmosphere (i.e., no flow) for 13 days and subsequently cooled to room temperature over a period of $\sim$18\,hr. Note that for all three bakes, the coupons were placed on a ceramic stage with their contaminated side placed face up.

\begin{figure}[t!]
\centering
\includegraphics[width=0.99\linewidth]{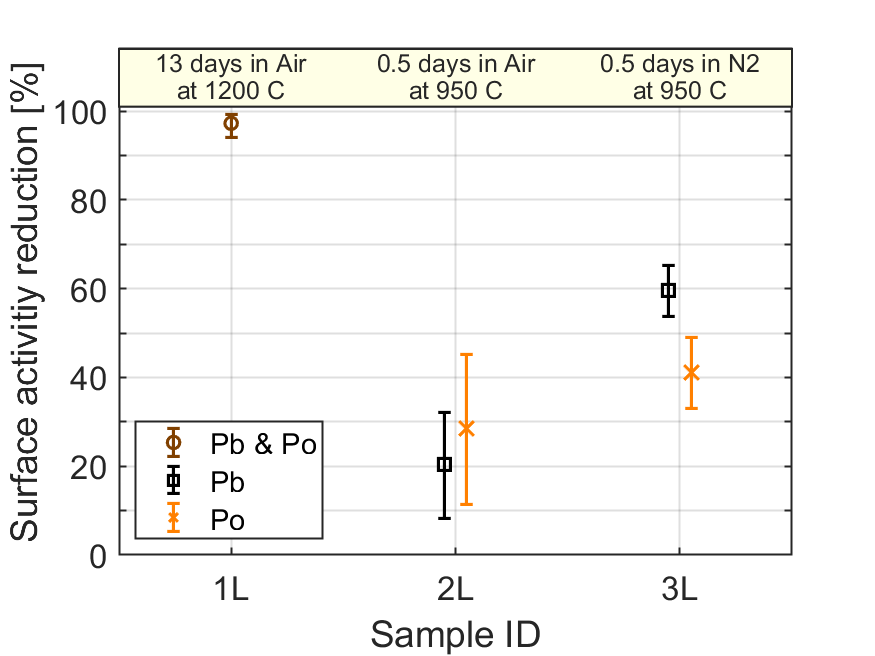}
\caption{Summary of baking results, showing the surface activity reductions for different parameters (indicated along the top) relative to the pre-treatment surface activities. For coupons 2L and 3L, reduction factors are determined for both $^{210}$Pb (black $\square$) and $^{210}$Po (orange $\times$), whereas for coupon 1L the data only permit estimation of a single, common reduction factor (brown $\circ$). The results for coupon 1L are consistent with complete removal from the coupon's surface and subsurface regions.}
\label{fig:heat}
\end{figure}

\begin{figure}[t!]
\centering
\includegraphics[width=0.99\linewidth]{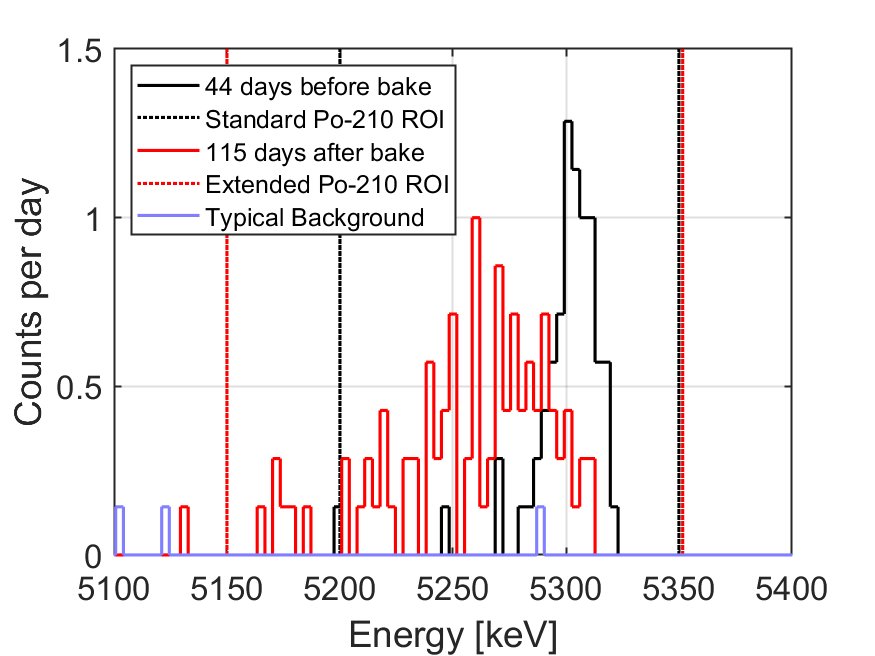}
\caption{Comparison of alpha spectra for coupon 3L before (black solid) and after (red solid) baking at 950 $^{\circ}$C in a nitrogen atmosphere for 12 hours. We attribute the broadening and softening of the post-bake spectrum to diffusion of the implanted $^{210}$Pb and $^{210}$Po to greater subsurface depths. To estimate the residual surface activity, the $^{210}$Po region of interest (ROI) was increased for analysis of post-bake coupons (red dotted) relative to the standard ROI (black dotted). A typical background spectrum is also shown (light blue). }
\label{fig:spectral_comp}
\end{figure}

Our heat-treatment results are summarized in Fig.~\ref{fig:heat}, which compares the best-fit surface activity reduction factors $f_{\mathrm{Pb}}$ and $f_{\mathrm{Po}}$ (from Eq.~\ref{eq:PoPost}) for the three coupons. The two lower-temperature bakes resulted in modest reductions and hint at a possible advantage when baking in pure nitrogen (relative to air).  Comparison of the pre- versus post-bake spectra for these two coupons (2L and 3L) shows a broadening and softening of the $^{210}$Po peak, which we attribute to diffusion of the remaining $^{210}$Pb and $^{210}$Po contamination to greater subsurface depths; the alphas must penetrate a greater thickness of silicon to be detected and are thus degraded in energy. Figure~\ref{fig:spectral_comp} shows the pre- versus post-bake spectral comparison for coupon 3L. To allow for the observed energy degradation, an extended ROI was used to estimate the post-treatment $^{210}$Po count rates for all three coupons. The more aggressive bake at 1200\,$^{\circ}$C resulted in  
spectra 
statistically consistent with the expected detector background. Only two such post-bake measurements were performed; so, the coupon 1L results in Fig.~\ref{fig:heat} correspond to a common $^{210}$Pb and $^{210}$Po reduction factor (see Sec.~\ref{ssec:assay}). Note that we cannot rule out that some of coupon 1L's pre-bake surface contamination diffused far enough into the silicon bulk such that alphas from any post-bake $^{210}$Po decays are unable to escape the coupon to be detected.

\subsection{Wet Chemistry}
\label{ssec:chemical}

As discussed in Sec.~\ref{sec:intro}, there have been many studies of methods utilizing wet chemistry to remove radon progeny from a variety of material surfaces.  However, there are surprisingly few measurements of $^{210}$Pb and $^{210}$Po removal from silicon surfaces. Considering the importance of silicon as a detector substrate, we therefore included wet-chemistry techniques in our study  both to address the apparent lack of data and as a point of reference. We explored several approaches, including wet etching, leaching and other solution-based treatments. 

In the case of leaching, the cleaning solution may partially remove  the surface oxide layer but leaves the underlying silicon-crystal structure in tact; surface contaminants are ``leached'' away from the surface via contact with (and thus dissolution into) the solution. Proton-donating acids such as nitric acid (HNO$_3$) are the most commonly used leaching solutions for removing metal cations from surfaces;  the H$^{+}$ cations, which are in high abundance in the leachate, can out-compete metal cations (e.g., $^{210}$Pb$^{2+}$ and $^{210}$Po$^{+}$) for the electron-rich binding sites near the surface (e.g., Si-O$^{-}$ and Si-O-O-Si). 

Cleaning via submersing a material in a solution (as with leaching) is an attractive possibility because it is perhaps the easiest method to apply (vs.\ etching, wiping, etc.), and when practical represents a mild treatment that will not alter physical properties such as material thickness and surface roughness.  \textit{A priori}, we did not expect that this approach would be fully effective for removal of radon progeny.  However, studies such as in Ref.~\cite{Stein:2017tel} have demonstrated that wiping can provide a significant reduction; we were curious if leaching or other similarly mild solution-based treatments could provide a comparable level of performance.

\begin{figure}[t!]
\centering
\includegraphics[width=0.99\linewidth]{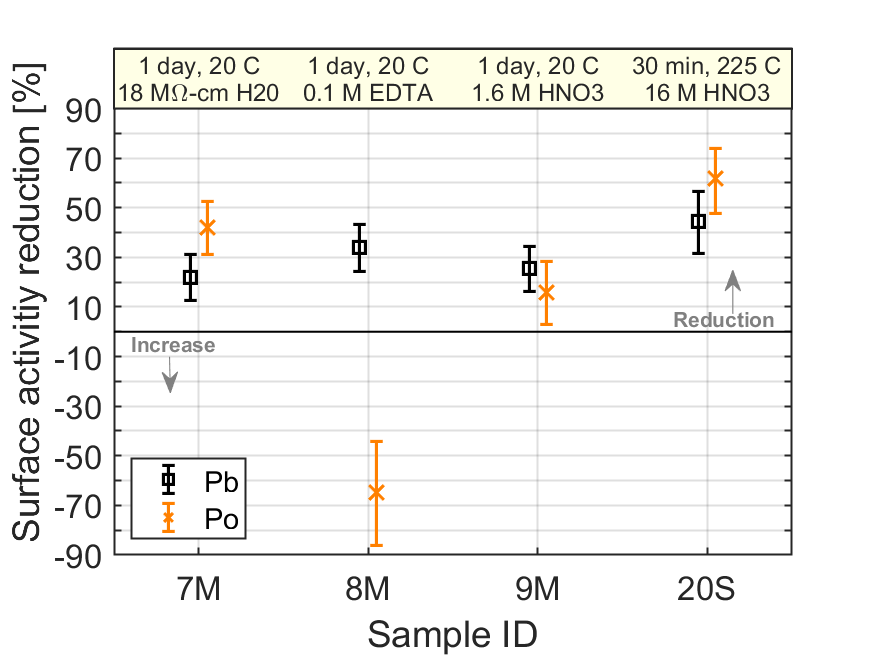}
\caption{Summary of leaching and other solution-based results, showing the reduction of $^{210}$Pb (black $\square$) and $^{210}$Po (orange $\times$) for different solutions and conditions (indicated along the top). Reduction factors at room temperature are modest and generally consistent, except for the anomalous increase in $^{210}$Po for coupon 8M following submersion in an EDTA solution. Using a stronger nitric-acid solution at elevated temperature on coupon 20S was most effective but still removed only $\sim$50\% of the surface contamination.}
\label{fig:leach}
\end{figure}

Toward this end, we cleaned three of the 2$\times$2\,cm$^2$ coupons at room temperature, each in a different solution:  high-purity water (H$_2$O) characterized by 18\,M$\Omega$-cm resistivity, a 0.1\,M solution of ethylenediaminetetraacetic acid (EDTA), and a 1.6\,M solution of HNO$_3$. Immediately after extraction from the cleaning solution, each coupon was rinsed three times with high-purity water and then allowed to air dry. As summarized in Fig.~\ref{fig:leach}, the $^{210}$Pb reduction factors at room temperature are modest and generally consistent with $\sim$30\% removal. The $^{210}$Po results are more surprising. Submersion in  water was the most effective of the three solutions, with  $\sim$40\% $^{210}$Po removal, whereas complexation with EDTA somehow increased the $^{210}$Po surface contamination.  We selected EDTA because of its capacity to attach to (and effectively trap) metal ions in solution. Our results are consistent with this occurring for $^{210}$Pb but appear to suggest that our EDTA was already carrying $^{210}$Po and thus increased the surface contamination on coupon 8M. We also considered if leaching efficacy could be enhanced by elevating the temperature and using a stronger solution. Coupon 20S was placed in a solution of 16\,M nitric acid and heated for $\sim$30\,min at 225\,$^{\circ}$C in a microwave digestion system. The resulting $^{210}$Pb and $^{210}$Po reduction factors are also shown in Fig.~\ref{fig:leach}.  Compared to the longer room-temperature leach in a weaker solution (coupon 9M), the elevated-temperature leaching of coupon 20S was more effective but still removed only $\sim$50\% of the surface contamination.

It is clear that a more aggressive treatment is required to fully remove implanted radon progeny from silicon surfaces using a wet-chemistry method.  We explored wet etching using hydrofluoric acid (HF). In this case, the silicon-solution chemistry enables both the oxide layer and some depth of the underlying crystal structure to be dissolved, depending on solution strength and duration of the etch. Solutions of HF were prepared to target different etch depths based on the stoichiometry of the HF-silicon reaction; each solution strength was determined by the target etch depth, the coupon's exposed surface area, and the volume of etching solution. To remove the surface oxide layer, the etching solutions were supplemented with nitric acid to a 1.5\,M concentration. Two 2$\times$2\,cm$^2$ coupons (2M and 6M) and one 3$\times$3\,cm$^2$ coupon (6L) were used.  The backsides of the two smaller coupons were masked with Kapton\textsuperscript{\textregistered} tape to restrict the etching to the contaminated surface, while coupon 6L was unmasked and thus the HF concentration was determined using its full surface area. Coupons 2M and 6M were submersed in 4\,mL solutions of 10 and 100\,mM HF, respectively, to approximately target 200 and 2000\,nm etch depths. The larger 6L coupon was submersed in a 10\,mL solution of 1.12\,mM HF to target an etch depth of 75\,nm.  The etch depths of 75 and 200\,nm were chosen to roughly bracket the maximum $^{210}$Pb implantation depth in silicon, while the much more aggressive 2000\,nm etch depth was chosen to ensure full removal of the surface contamination. In all cases, the coupons were left submersed for several days to allow sufficent time for the HF-silicon reaction to go to completion:  6 days each for coupons 2M and 6M, and 14 days for coupon 6L. Following the etch, each coupon was rinsed three times with high-purity water and allowed to air dry. 

\begin{figure}[t!]
\centering
\includegraphics[width=0.99\linewidth]{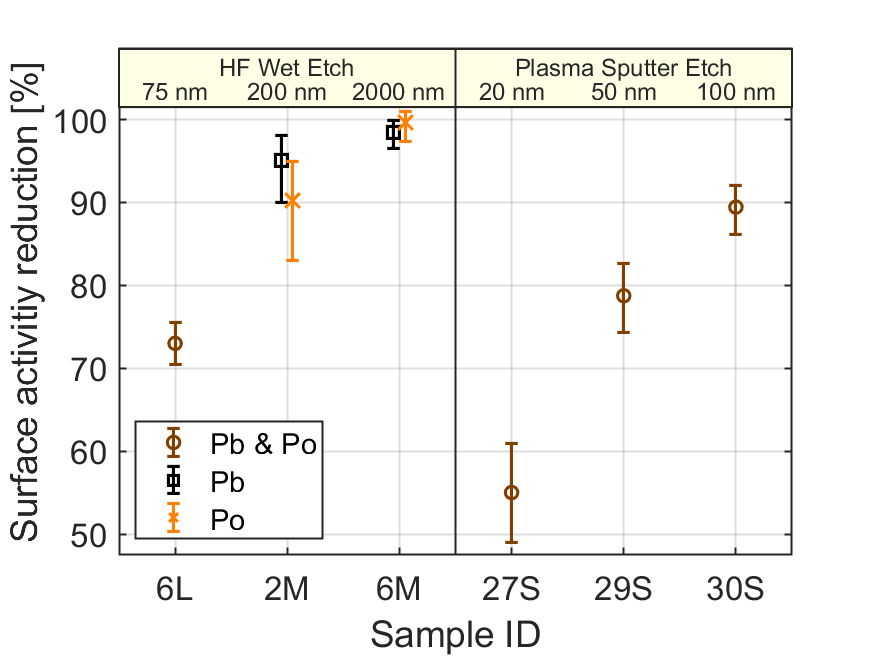}
\caption{Summary of etching results, comparing the surface activity reductions for HF wet etching (left) vs.\ plasma sputter etching (right) for different target depths (indicated along the top). Although the radon progeny are implanted up to $\mathcal{O}$(100\,nm) depth, a relatively aggressive etch appears required to achieve 100\% reduction. For coupons 2M and 6M, reductions factors are determined for both $^{210}$Pb (black $\square$) and $^{210}$Po (orange $\times$), whereas for 6L, 27S, 29S, and 30S the data only permit estimation of a single, common reduction factor (brown $\circ$).}
\label{fig:etch}
\end{figure}

Our wet-etching results are summarized in the left panel of Fig.~\ref{fig:etch}, which compares the best-fit surface activity reduction factors $f_{\mathrm{Pb}}$ and $f_{\mathrm{Po}}$ for the three coupons. The target etch depths are indicated along the top. Only two post-etch measurements were performed for coupon 6L and thus a common $^{210}$Pb and $^{210}$Po reduction factor was assumed in its analysis. Qualitatively, we observe the expected trend of increasing reduction factor with increased etch depth. 
We also note that while the 200\,nm etch was effective---as expected based on the $^{210}$Pb implantation depth---it was not sufficient for complete removal of the surface contamination. The results of the more aggressive 2000\,nm etch are consistent with full removal.  

\subsection{Plasma Etching}
\label{ssec:plasma}
We also explored use of a plasma to sputter etch silicon surfaces. Like baking, plasma etching is of particular interest for solid-state detectors because it is often used as a processing step during sensor fabrication (see, e.g., use of reactive ion etching in Ref.~\cite{Jastram:2014pga}). It is therefore of interest to explore the efficacy of plasma etching for mitigating surface contamination resulting from any exposure to radon that occurred prior to plasma-etch processing steps during device fabrication.

The sputter etching was performed in an ultrahigh-vacuum chamber (base pressure 10$^{-10}$ Torr) using a ULVAC-PHI focused beam ion gun. Ne$^{+}$ ions were accelerated using a 2.0\,kV bias (20\,mA emission current) and rastered over the sample area. We selected three of the small 1$\times$1\,cm$^2$ coupons to subject to this plasma-based etching method.  For removal of 20, 50, and 100\,nm, etching times of 0.5, 1.25, and 2.5 hours were used. These etching times were calculated based on a calibration run in which a 12 hour scan resulted in a 480\,nm etch depth. We did not independently assess the etch depths achieved for all three of the 1$\times$1\,cm$^2$. 
However, optical profilometry of coupon 27S revealed an $\sim$20\,nm feature associated with a small wire used to hold the coupon in place during the etch, which gives us confidence that the sputter-etch depth scales linearly as a function of etch duration.

Our sputter-etching results are summarized in the right panel of Fig.~\ref{fig:etch}. Because of the smaller size of these 1$\times$1\,cm$^2$ coupons, the statistical precision of the post-etch alpha measurements is relatively poor. As a result, despite making several post-etch measurements over a period of $\sim$6 months, the data are insufficient to measure a difference between  $f_{\mathrm{Pb}}$ and $f_{\mathrm{Po}}$. We therefore assumed a common reduction factor in the analysis of all three coupons. Similar to the wet-etching results, we observe the expected dependence of the reduction factor on the etch depth, which increases with increasing etch depth. The 100\,nm etch was highly effective with $\sim$90\% of the surface contamination removed; a deeper etch appears to be required for complete removal.  Our results also suggest that Ne$^{+}$ sputter etching is more effective (versus etch depth) than the HF-based wet etching for removal of radon progeny from silicon surfaces. 

\section{Summary}
\label{sec:summary}
In this article, we document our exploration of a variety of mitigation methods to reduce a source of background events that is a key concern for many rare-event searches: implanted $^{210}$Pb surface contamination resulting from radon exposure. Considering the many materials for which such studies are already reported in the literature, we focused our efforts on silicon because it is an important substrate in several experiments~\cite{Agnese:2016cpb,DAMIC:2020cut,SENSEI:2020dpa,Aguilar-Arevalo:2022kqd,CDMS:2013juh,SuperCDMS:2020ymb,SuperCDMS:2020aus} and there are surprisingly few prior measurements of the type reported here. Silicon coupons implanted with $^{210}$Pb and $^{210}$Po were created via exposure to radon and then subjected to various treatments to test their efficacy for removal of the radioactive contaminants. Our study is unique in its exploration of a broad range of chemical and physical processes, including heat treatments, wet chemistry, and sputter etching. The silicon coupons were assayed with an alpha detector both before and after application of the surface treatments to measure the $^{210}$Po surface activity as a function of time, enabling estimation of the $^{210}$Pb and $^{210}$Po reduction factors and thus assessment of treatment effectiveness.

Generally, all of the methods we explored are at least partially effective for removal of implanted radon progeny, with a few consistent with complete removal.  We conclude that baking has potential as a low-background cleaning method, and we note that it may be applicable to a wide range of materials, radioactive contaminants, and fabrication processes. For example, baking to reduce near-surface $^{226}$Ra may be a viable option for mitigation of radon emanation. To our knowledge, our results represent the first demonstration that implanted radon progeny can be removed via baking. Although leaching is only partially effective, it is notable that submersing a material in high-purity water is a straightforward cleaning method that results in a nontrivial reduction of the $^{210}$Pb and $^{210}$Po surface activities. Our results show that wet etching silicon in a solution of HF and HNO$_3$ can be highly effective. However, complete removal of the implanted surface contamination appears to require a larger etch depth than na\"ively expected based on the stoichiometry of the HF-silicon reaction and $^{210}$Pb implantation depth.  Similarly, sputter etching of silicon surfaces using a plasma is highly effective; even a modest 20\,nm etch removed half of the radioactive surface contaminants. Relative to the other cleaning methods, plasma-based etching is limited in its applicability because of the equipment required.  However, sensor fabrication for solid-state detectors often includes plasma-etch steps.  Even a light cleaning of the substrate surface (e.g., as described in Ref.~\cite{Jastram:2014pga}) likely results in a useful reduction of any radon-progeny surface contamination that may have accumulated up to that point.

In summary, our results demonstrate the utility of a wide range of methods for removal of implanted radon progeny from material surfaces. We emphasize that several of these cleaning methods may be applicable in the fabrication and assembly of low-background detectors, in particular for silicon-based devices but for other detector materials as well (e.g., fused silica/quartz, sapphire, stainless steel, and titanium). Further, fabrication procedures may already involve use of elevated temperatures, wet chemistry, and/or plasma beams; our results suggest the potential to optimize such steps for convenient mitigation of surface contaminants. Finally, following recent developments in the literature regarding the influence of radiation on superconducting circuits~\cite{Vepsalainen:2020trd,Cardani:2020vvp,Wilen:2020lgg}, we note that cleaning of radioactive surface contaminants may become an important consideration as the sensitivity of these quantum devices continues to improve.

\section*{Acknowledgments}
We gratefully acknowledge technical assistance from several staff at the Pacific Northwest National Laboratory (PNNL):  Shannon Morley for assistance with preparation of the silicon coupons, Brian Glasgow and Grant Spitler for operation of the alpha counters, Derek Cutforth for baking of coupon 1L in the vertical muffle furnace, and Ben Loer for useful feedback on this manuscript.

This work was funded by PNNL Laboratory Directed Research
and Development (LDRD) funds under the Nuclear Physics, Particle Physics, Astrophysics, and Cosmology Initiative. For the plasma etching, Zdenek Dohnalek also acknowledges the PNNL Chemical Dynamics LDRD initiative. PNNL is a multi-program national laboratory operated for the U.S.\ Department of Energy (DOE) by Battelle Memorial Institute under contract number DE-AC05-76RL01830.

\bibliography{mybib} 
\end{document}